\documentclass{article}

\usepackage{graphicx,float}
\usepackage{t1enc}
\usepackage{epsfig}
\usepackage[francais,]{babel}
\usepackage{amssymb,amsbsy,amsmath,amsfonts,amssymb,amscd}

\newtheorem{theoreme}{Th\'eor\`eme}[section]

\newtheorem{proposition}[theoreme]{Proposition}
\newtheorem{corollaire}[theoreme]{Corollaire}

\newtheorem{remarque}{\it Remarque}

\begin{document}
\title{\bf {\Huge Critique \\ du rapport
signal \`{a} bruit en th\'{e}orie de l'information} \\  {\normalsize {\em A
critical appraisal of the signal to noise ratio in information
theory}}}  \vspace{3cm}
\author{Michel FLIESS \\   {\small{ Projet ALIEN, INRIA Futurs}} \\
\small{ \& \'Equipe MAX, LIX (CNRS, UMR 7161)}
\\  \small{\'{E}cole polytechnique, 91128 Palaiseau, France} \\
\small{ E-mail: {\tt Michel.Fliess@polytechnique.edu}}}
\date{}


\maketitle

\small{\noindent{\bf R\'{e}sum\'{e}}. On d\'{e}montre que le rapport signal \`{a}
bruit, si important en th\'{e}orie de l'information, devient sans objet
pour des communications num\'{e}riques o\`{u}
\begin{itemize}
\item les symboles modulent des porteuses, solutions d'\'{e}quations
diff\'{e}rentielles lin\'{e}aires \`{a} coefficients polynomiaux;
\item de nouvelles techniques alg\'{e}briques
d'estimation permettent la d\'{e}modulation. \end{itemize} Calcul
op\'{e}rationnel, alg\`{e}bre diff\'{e}rentielle et
analyse non standard sont les principaux outils math\'{e}matiques.  \\

\noindent{\bf Abstract}. The signal to noise ratio, which plays such
an important role in information theory, is shown to become
pointless in digital communications where \begin{itemize}
\item symbols are modulating carriers, which are solutions of linear
differential equations with polynomial coefficients,
\item demodulations is achieved thanks to new algebraic estimation
techniques.
\end{itemize}Operational calculus, differential algebra and nonstandard
analysis are the main mathematical tools. \\


}

\newpage

\section{Introduction}
\label{intro} Le {\em rapport signal/bruit}, que l'on retrouve dans
les formules de la th\'{e}orie de l'information, telle qu'elle s'est
impos\'{e}e depuis Shannon (voir \cite{bell,ire} et, par exemple, dans
la vaste litt\'{e}rature sur le sujet,
\cite{battail,blahut,brillouin,cover,proakis}), est un ingr\'{e}dient
fondamental pour d\'{e}finir la qualit\'{e} des communications. Le but de
cette Note est de d\'{e}montrer qu'une nouvelle approche de l'estimation
rapide et du bruit (voir \cite{ans} et sa bibliographie) rend ce
rapport sans objet dans un certain cadre num\'{e}rique. Revoyons, donc,
le \og paradigme de Shannon \fg. Le symbole \`{a} transmettre (voir, par
exemple, \cite{glavieux,proakis}) module une porteuse $z(t)$,
solution d'une \'{e}quation diff\'{e}rentielle lin\'{e}aire \`{a} coefficients
polynomiaux $\sum_{\tiny{\mbox{\rm finie}}} a_\nu (t) z^{(\nu)} (t)
= 0$, $a_\nu \in \mathbb{C}[t]$. La plupart des signaux utilis\'{e}s en
pratique, comme une somme trigonom\'{e}trique finie
$\sum_{\tiny{\mbox{\rm finie}}} A_\iota \sin (\omega_\iota t +
\varphi_\iota)$, un sinus cardinal $\frac{\sin (\omega t)}{t}$ ou un
cosinus sur\'{e}lev\'{e} $\frac{\cos (\omega t)}{1 + t^2}$, $A_\iota,
\omega_\iota, \varphi_\iota, \omega \in \mathbb{R}$, v\'{e}rifient une
telle \'{e}quation, qui se traduit dans le domaine op\'{e}rationnel (cf.
\cite{yosida}), pour $t \geq 0$, par
\begin{equation}\label{nh0} \sum_{\tiny{\mbox{\rm finie}}} a_\nu (-
\frac{d}{ds}) s^\nu \hat{z} = I(s) \end{equation} o\`{u} $I \in
\mathbb{C}[s]$ est un polyn\^{o}me dont les coefficients d\'{e}pendent des
conditions initiales en $t = 0$. La d\'{e}modulation revient, alors, \`{a}
estimer certains des coefficients de (\ref{nh0}). On y parvient,
ici, gr\^{a}ce \`{a} des techniques alg\'{e}briques r\'{e}centes (cf.
\cite{esaim,garnier}).

Un bruit, selon \cite{ans}, est une fluctuation rapide, d\'{e}finie dans
le langage de l'analyse non standard, que nous ne rappellerons pas
ici (voir aussi \cite{lobry} et sa bibliographie).
Les calculs du {\S} \ref{bruit} sont effectu\'{e}s avec une somme finie de
sinuso\"{\i}des \`{a} tr\`{e}s hautes fr\'{e}quences et un bruit blanc, dont la
d\'{e}finition non standard, \`{a} comparer avec celle de \cite{al},
clarifie l'approche usuelle des manuels de traitement du signal. Ils
d\'{e}montrent la possibilit\'{e} d'obtenir de \og bonnes \fg ~ estimations
avec des bruits \og tr\`{e}s forts \fg, c'est-\`{a}-dire de \og grandes \fg
~ puissances, fait confirm\'{e} par des simulations num\'{e}riques et des
expriences de laboratoire (voir, par exemple,
\cite{delaleau,mboup,neves,trapero,trapero-bis} et certaines
r\'{e}f\'{e}rences de \cite{ans}). Les imperfections, in\'{e}vitables en
pratique, proviennent de l'implantation num\'{e}rique des calculs,
notamment de celui des int\'{e}grales (cf. \cite{mboup,ath}), des
interf\'{e}rences entre symboles (voir
\cite{battail,glavieux,proakis0,proakis} et leur bibliographie), et
du fait que les bruits ne sont pas n\'{e}cessairement centr\'{e}s (voir \`{a} ce
propos le {\S} 3.2.2 de \cite{ans}).

Calcul op\'{e}rationnel et alg\`{e}bre diff\'{e}rentielle aux {\S} \ref{algebre} et
\ref{perturb}, analyse non standard au {\S} \ref{bruit} sont les
principaux outils math\'{e}matiques.

\begin{remarque}
Avec des signaux analytiques par morceaux (le sens du mot {\em
analytique} est celui de la th\'{e}orie des fonctions et non pas, ici,
celui usuel en traitement du signal (cf.
\cite{battail,proakis0,proakis})), qui ne satisfont pas d'\'{e}quations
diff\'{e}rentielles connues \`{a} l'avance, on utilise, selon les m\^{e}mes
principes alg\'{e}briques, des d\'{e}rivateurs num\'{e}riques \`{a} fen\^{e}tres
glissantes pour obtenir les estimations (voir \cite{ath}, \cite{nl},
leurs exemples et bibliographies). On ne peut, alors, esp\'{e}rer les
m\^{e}mes r\'{e}sultats que pr\'{e}c\'{e}demment.
\end{remarque}

\begin{remarque}
La possibilit\'{e} de liens entre th\'{e}orie de l'information et m\'{e}canique
quantique a \'{e}t\'{e} examin\'{e}e par divers auteurs (voir, par exemple,
\cite{brillouin,austria,green}). Rappelons \`{a} ce propos que
l'approche du bruit en \cite{ans} a d\'{e}j\`{a} conduit \`{a} une tentative
nouvelle de formalisation du quantique \cite{mecaqua}.
\end{remarque}

\section{Identifiabilit\'{e}}\label{algebre}
\subsection{\'Equations diff\'{e}rentielles}
Renvoyons \`{a} \cite{chambert} pour des rappels sur les corps,
diff\'{e}rentiels ou non. Soit $k_0$ le corps de base de caract\'{e}ristique
nulle, $\mathbb{Q}$ par exemple. Soit $k_0 (\Theta)$ le corps
engendr\'{e} par un ensemble fini $\Theta = \{\theta_1, \dots,
\theta_\varrho \}$ de {\em param\`{e}tres} inconnus. Soit $\bar{k}$ la
cl\^{o}ture alg\'{e}brique de $k_0 (\Theta)$. Introduisons le corps
$\bar{k}(s)$ des fractions rationnelles en l'ind\'{e}termin\'{e}e $s$, que
l'on munit d'une structure de corps diff\'{e}rentiel gr\^{a}ce \`{a} la
d\'{e}rivation $\frac{d}{ds}$ (les \'{e}l\'{e}ments de $k_0$, de $\Theta$ et,
donc, de $\bar{k}$, sont des constantes). Tout {\em signal} $x$, $x
\not\equiv 0$, est suppos\'{e} satisfaire une \'{e}quation diff\'{e}rentielle
lin\'{e}aire homog\`{e}ne, \`{a} coefficients dans $\bar{k}(s)$, et donc
appartenir \`{a} une extension de Picard-Vessiot de $\bar{k}(s)$.

\begin{remarque}
Il suffit pour se convaincre de l'existence d'une telle \'{e}quation
homog\`{e}ne de d\'{e}river les deux membres de (\ref{nh0}) suffisamment de
fois par rapport \`{a} $s$.
\end{remarque}

L'anneau non commutatif $\bar{k}(s) [\frac{d}{ds}]$ des op\'{e}rateurs
diff\'{e}rentiels lin\'{e}aires \`{a} coefficients dans $\bar{k}(s)$ est
principal \`{a} droite et \`{a} gauche. Le $\bar{k}(s)
[\frac{d}{ds}]$-module \`{a} gauche engendr\'{e} par $x$ et $1$ est un
module de torsion (cf. \cite{McC}), et, donc, un $\bar{k}(s)$-espace
vectoriel de dimension finie, $n + 1$, $n \geq 0$. D'o\`{u} le r\'{e}sultat
suivant qui semble nouveau (cf. \cite{chambert,singer}):

\begin{proposition}
Il existe un entier minimal $n \geq 0$, tel que $x$ satisfait
l'\'{e}quation diff\'{e}rentielle lin\'{e}aire, d'ordre $n$, non n\'{e}cessairement
homog\`{e}ne,
\begin{equation}\label{nh}
\left(\sum_{\iota = 0}^{n} q_\iota \frac{d^\iota}{ds^\iota}\right) x
- p = 0
\end{equation}
o\`{u} les polyn\^{o}mes $p, q_0, \dots, q_{n} \in \bar{k}[s]$ sont premiers
entre eux. Cette \'{e}quation, dite {\em minimale}, est unique \`{a} un
coefficient multiplicatif constant non nul pr\`{e}s.
\end{proposition}

\subsection{Identifiabilit\'{e} lin\'{e}aire projective}
Rappelons que l'ensemble $\Theta = \{\theta_1, \dots, \theta_\varrho
\}$ de param\`{e}tres est dit (cf. \cite{esaim,garnier})
\begin{itemize}
\item {\em lin\'{e}airement identifiable} si, et selement si,
\begin{equation}\label{li}
\mathfrak{A} \left(\begin{array}{c} \theta_1 \\ \vdots \\
\theta_\varrho \end{array} \right) = \mathfrak{B}
\end{equation}
o\`{u}
\begin{itemize}
\item les entr\'{e}es des matrices $\mathfrak{A}$, carr\'{e}e $\varrho
\times \varrho$, et $\mathfrak{B}$, colonne $\varrho \times 1$,
appartiennent \`{a} $\mbox{\rm span}_{k_0 (s) [\frac{d}{ds}]} (1, x)$;
\item $\det ( \mathfrak{A}) \neq 0$.
\end{itemize}
\item {\em projectivement lin\'{e}airement identifiable} si, et
seulement si,
\begin{itemize}
\item il existe un param\`{e}tre, $\theta_1$ par exemple, non nul,
\item l'ensemble $\{ \frac{\theta_2}{\theta_1}, \dots,
\frac{\theta_\varrho}{\theta_1} \}$ est lin\'{e}airement identifiable.
\end{itemize}
\end{itemize}
R\'{e}\'{e}crivons (\ref{nh}) sous la forme suivante:
\begin{equation}\label{nhcoef}
\left( \sum_{\tiny{\mbox{\rm finie}}} a_{\mu \nu} s^\mu
\frac{d^\nu}{ds^\nu} \right) x - \sum_{\tiny{\mbox{\rm finie}}}
b_\kappa s^\kappa = 0
\end{equation}
o\`{u} les $N + 1$ coefficients $a_{\mu \nu}$ et les $M$ coefficients
$b_\kappa$ appartiennent \`{a} $\bar{k}$. La matrice carr\'{e}e
$\mathfrak{M}$ d'ordre $N + M + 1$, dont la $\xi^{\tiny\mbox{\rm
\`{e}me}}$ ligne, $0 \leq \xi \leq N + M$, est
\begin{equation*}\label{line} \dots,
\frac{d^\xi}{ds^\xi} \left( s^\mu \frac{d^\nu x}{ds^\nu}\right),
\dots, \frac{d^\xi s^\kappa}{ds^\xi}, \dots
\end{equation*}
est singuli\`{e}re d'apr\`{e}s (\ref{nh}) et (\ref{nhcoef}). La minimalit\'{e}
de (\ref{nh}) permet de d\'{e}montrer selon des techniques bien connues
sur le rang du wronskien (cf. \cite{chambert,singer}) que le rang de
$\mathfrak{M}$ est $N + M$. Il en d\'{e}coule:
\begin{theoreme}
Les coefficients $a_{\mu \nu}$ et $b_\kappa$ de (\ref{nhcoef}) sont
projectivement lin\'{e}airement identifiables.
\end{theoreme}
\begin{corollaire}
Posons $x = \frac{p(s)}{q(s)}$, o\`{u} les polyn\^{o}mes $p, q \in \bar{k}
[s]$ sont premiers entre eux. Alors, les coefficients de $p$ et $q$
sont projectivement lin\'{e}airement identifiables.
\end{corollaire}
Il est loisible de supposer l'ensemble des param\`{e}tres inconnus
$\Theta = \{\theta_1, \dots, \theta_\varrho \}$ strictement inclus
dans celui des coefficients $a_{\mu \nu}$ et $b_\kappa$ de
(\ref{nhcoef}), et donc lin\'{e}airement identifiable.

\section{Perturbations et estimateurs}\label{perturb}
Avec une perturbation additive $w$ le capteur fournit non pas $x$
mais $x + w$. Soient $R =  k_0(\Theta)[s] (k_0 [s])^{-1}$ l'anneau
{\em localis\'{e}} (cf. \cite{lang}) des fractions rationnelles \`{a}
num\'{e}rateurs dans $k_0(\Theta)[s]$ et d\'{e}nominteurs dans $k_0 [s]$, et
$R[\frac{d}{ds}]$ l'anneau non commutatif des op\'{e}rateurs
diff\'{e}rentiels lin\'{e}aires \`{a} coefficients dans $R$. On obtient, \`{a}
partir de (\ref{li}), la
\begin{proposition}\label{estimperturb}
Les param\`{e}tres inconnus v\'{e}rifient
\begin{equation}\label{estim}
\mathfrak{A} \left(\begin{array}{c} \theta_1 \\ \vdots \\
\theta_\varrho \end{array} \right) = \mathfrak{B} + \mathfrak{C}
\end{equation}
o\`{u} les entr\'{e}es de $\mathfrak{C}$, matrice colonne $\varrho \times
1$, appartiennent \`{a} $\mbox{\rm span}_{R [\frac{d}{ds}]} (w)$.
\end{proposition}
On appelle (\ref{estim}) un {\em estimateur}. Il est dit {\em
strictement polynomial en $\frac{1}{s}$} si, et seulement si, toutes
les fractions rationnelles en $s$, rencontr\'{e}es dans les coefficients
des matrices $\mathfrak{A}$, $\mathfrak{B}$, $\mathfrak{C}$ de
(\ref{estim}), sont des polyn\^{o}mes en $\frac{1}{s}$ sans termes
constants. On peut toujours s'y ramener en multipliant les deux
membres de (\ref{estim}) par une fraction rationnelle de $k_0(s)$
convenable. On aboutit, alors, dans le domaine temporel, aux
estimateurs consid\'{e}r\'{e}s en \cite{ans}, si l'on suppose l'analyticit\'{e}
du signal:

\begin{equation}\label{estimat}
\delta  (t) \left( [ \theta_\iota ]_e (t) - \theta_\iota \right) =
\sum_{\tiny{\mbox{\rm finie}}} c \int_{0}^{t} \dots
\int_{0}^{\tau_2} \int_{0}^{\tau_1} \tau_{1}^{\nu} w(\tau_1)d\tau_1
d\tau_2 \dots d\tau_k  \quad \quad ~ ~\iota = 1, \dots, \varrho
\end{equation}
o\`u
\begin{itemize}
\item $c$ est une constante,
\item $[0, t]$ est la {\em fen\^etre d'estimation}, de {\em largeur}
$t$,
\item $\delta (t)$ est une fonction analytique,
appel\'{e}e {\em diviseur}, nulle en $0$,
\item $[ \theta ]_e (t)$ est l'estim\'{e}e de $\theta$ en $t$.
\end{itemize}

\section{Bruits}\label{bruit}
Renvoyons \`{a} \cite{robinson} et \cite{diener} pour la terminologie de
l'analyse non standard, d\'{e}j\`{a} utilis\'{e}e en \cite{ans}. On trouvera une
excellente introduction \`{a} cette analyse en \cite{lobry}. Les
propositions \ref{propsin} et \ref{propbr} ci-dessous affinent la
proposition 3.2 de \cite{ans}, o\`{u} les estimations sont obtenues en
temps fini, court en pratique.

\subsection{Sinuso\"{\i}des hautes fr\'{e}quences}\label{sinus}
La perturbation du {\S} \ref{perturb} est une somme finie
$\sum_{\tiny{\mbox{\rm finie}}} A_\iota \sin (\Omega_\iota t +
\varphi_\iota)$ de sinuso\"{\i}des, dont les fr\'{e}quences $\Omega_\iota
>0$ sont illimit\'{e}es: c'est un bruit centr\'{e} au sens de
\cite{ans}. Des manipulations \'{e}l\'{e}mentaires des int\'{e}grales it\'{e}r\'{e}es
(\ref{estimat}) conduisent \`{a} la

\begin{proposition}\label{propsin} Si \begin{itemize}
\item les quotients $\frac{A_\iota}{\Omega_\iota}$ sont infinit\'{e}simaux,

\item la largeur de la fen\^{e}tre d'estimation est limit\'{e}e et n'appartient
pas au halo d'un z\'{e}ro du diviseur,
\end{itemize}
les estim\'{e}es des param\`{e}tres inconnus, obtenues gr\^{a}ce \`{a}
(\ref{estimat}), appartiennent aux halos de leurs vraies valeurs. Il
n'en va plus de m\^{e}me si l'un des quotients
$\frac{A_\iota}{\Omega_\iota}$ est appr\'{e}ciable.
\end{proposition}
\begin{corollaire}
Il existe des valeurs illimit\'{e}es des amplitudes $A_\iota$,
$\sqrt{\Omega_\iota}$ par exemple, telles que les estim\'{e}es
pr\'{e}c\'{e}dentes appartiennent aux halos des vraies valeurs.
\end{corollaire}

\subsection{Bruits blancs}
D\'{e}signons par $^*\mathbb{N}$, $^*\mathbb{R}$ les extensions non
standard de $\mathbb{N}$, $\mathbb{R}$. Rempla\c{c}ons l'intervalle $[0,
1] \subset {\mathbb{R}}$ par l'ensemble hyperfini ${\mathrm{I}} =
\{0, \frac{1}{\bar{N}}, \dots, \frac{\bar{N} - 1}{\bar{N}}, 1 \}$,
o\`u $\bar{N} \in {^*\mathbb{N}}$ est illimit\'{e}. Un {\em bruit blanc
centr\'{e}} est une fonction $w: {\mathrm{I}} \rightarrow
{^*\mathbb{R}}$, $\iota \mapsto w(\iota) = A n(\iota)$, o\`{u}
\begin{itemize}
\item $A \in {^*\mathbb{R}}$, $A
> 0$, est constant,
\item les $n(\iota)$ sont des variables al\'{e}atoires
r\'{e}elles, suppos\'{e}es centr\'{e}es, de m\^{e}me \'{e}cart-type $1$ normalis\'{e}, et
deux \`{a} deux ind\'{e}pendantes.
\end{itemize}
\begin{remarque}
Cette d\'{e}finition, qui pr\'{e}cise \cite{ans}, est inspir\'{e}e de
publications d'ing\'{e}nieurs sur le bruit blanc en temps discret (voir,
par exemple, \cite{proakis0}). Elle simplifie, \`{a} la mani\`{e}re de
\cite{nelson}, l'approche en temps continu usuelle dans les manuels
de traitement du signal (voir, \`{a} ce sujet,
\cite{battail,cover,proakis0,proakis} et leurs bibliographies).
Rappelons que cette approche continue est bas\'{e}e, en g\'{e}n\'{e}ral, sur
l'analyse de Fourier et renvoyons, \`{a} ce sujet, \`{a} \cite{fourier}.
\end{remarque}

Comme au {\S} \ref{sinus}, il vient:

\begin{proposition}\label{propbr} Si \begin{itemize}
\item le quotient $\frac{A}{\bar{N}}$ est infinit\'{e}simal,

\item la largeur $t$, $t \in {\mathrm{I}}$, de la fen\^{e}tre d'estimation
n'appartient pas au halo d'un z\'{e}ro du diviseur,
\end{itemize}
les estim\'{e}es des param\`{e}tres inconnus, obtenues gr\^{a}ce \`{a}
(\ref{estimat}), appartiennent presque s\^{u}rement aux halos de leurs
vraies valeurs. Il n'en va plus de m\^{e}me si le quotient
$\frac{A}{\bar{N}}$ est appr\'{e}ciable.
\end{proposition}
\begin{corollaire}
Il existe des valeurs illimit\'{e}es de $A$, $\sqrt{\bar{N}}$ par
exemple, telles que les estim\'{e}es pr\'{e}c\'{e}dentes appartiennent presque
s\^{u}rement aux halos des vraies valeurs.
\end{corollaire}

\begin{remarque}
Il est loisible de remplacer l'ind\'{e}pendance de $n(\iota)$ et
$n(\iota^\prime)$, $\iota \neq \iota^\prime$, par le fait que
l'esp\'{e}rance du produit $n(\iota) n(\iota^\prime)$ est
infinit\'{e}simale.
\end{remarque}

\vspace{0.2cm} \noindent{\bf Remerciements}. L'auteur exprime sa
reconnaissace \`{a} O. Gibaru (Lille), M. Mboup (Paris) et \`{a} tous les
membres du projet ALIEN, de l'INRIA Futurs, pour des \'{e}changes
fructueux.





\begin{thebibliography}{99}

\bibitem{al}S. Albeverio, J.E. Fenstad, R. Hoegh-Kr{\o}hn, T. Lindstr{\o}m,
Nonstandard Methods in Stochastic Analysis and Mathematical Physics,
Academic Press, Orlando, FL, 1986.
\bibitem{battail}G. Battail, Th\'{e}orie de l'information - Application
aux techniques de communication, Masson, Paris, 1997.

\bibitem{blahut}R.E. Blahut, Principles and Practice of Information Theory,
Addison-Wesley, Reading, MA, 1987.

\bibitem{delaleau}J.-M. Bourgeot, E. Delaleau, Fast algebraic impact times
estimation for a linear system subject to unilateral constraint,
Proc. 46$^{th}$ IEEE Conf. Decision Control - CDC 2007, New Orleans,
2007.


\bibitem{brillouin}L. Brillouin, Science and Information Theory, 2$^{nd}$ ed., Academic
Press, New York, 1962. Traduction fran\c{c}aise de la 1$^{re}$ \'{e}d.: La
science et la th\'{e}orie de l'information, Masson, Paris, 1959.

\bibitem{austria}C. Brukner, A. Zeilinger, Conceptual inadequacy of the Shannon information
in quantum measurements, Phys. Rev. A, 63 (2001) 022113.


\bibitem{cover}T.M. Cover, J.A. Thomas, Elements of Information Theory,
Wiley, New York, 1991.

\bibitem{chambert} A. Chambert-Loir, Alg\`{e}bre corporelle, \'Editions \'Ecole
Polytechnique, Palaiseau, 2005. English translation$:$  A Field
Guide to Algebra, Springer, Berlin, 2005.

\bibitem{diener}F. Diener, G. Reeb, Analyse non standard, Hermann,
Paris, 1989.


\bibitem{fourier}M. Fliess, R\'{e}flexions sur la question fr\'{e}quentielle en
traitement du signal, Manuscrit, 2005. Accessible sur {\tt
http$:$//hal.inria.fr/inria-00000461}.

\bibitem{ans}M. Fliess, Analyse non standard du bruit,
C.R. Acad. Sci. Paris Ser. I, 342 (2006) 797-802.

\bibitem{mecaqua}M. Fliess, Probabilit\'{e}s et fluctuations quantiques,
C.R. Acad. Sci. Paris Ser. I, 344 (2007) 663-668.


\bibitem{nl}M. Fliess, C. Join, H. Sira-Ram\'{\i}rez, Non-linear estimation
is easy, Int. J. Modelling Identification Control, 3 (2008).
Accessible sur {\tt http$:$//hal.inria.fr/inria-00158855}.

\bibitem{esaim}M. Fliess, H. Sira-Ram\'{\i}rez, An algebraic
framework for linear identification, ESAIM Control Optim. Calc.
Variat., 9 (2003) 151-168.

\bibitem{garnier}M. Fliess, H. Sira-Ram\'{\i}rez, Closed-loop parametric
identification for continuous-time linear systems via new algebraic
techniques, in H. Garnier \& L. Wang (Eds)$:$ Continuous-Time Model
Identification from Sampled Data, Springer, Berlin, 2008. Accessible
sur {\tt http$:$//hal.inria.fr/inria-00114958}.

\bibitem{green}H.S. Green, Information Theory and Quantum Physics,
Springer, Berlin, 2000.

\bibitem{glavieux}M. Joindot, A. Glavieux, Introduction aux communications num\'{e}riques,
Masson, Paris, 1996.

\bibitem{lang}S. Lang, Algebra, 3$^{rd}$ rev. ed., Springer, Berlin, 2002.
Traduction fran\c{c}aise: Alg\`ebre, Dunod, Paris, 2004.

\bibitem{lobry}C. Lobry, T. Sari, Nonstandard analysis and representation
of reality, Internat. J. Control, \`{a} para\^{\i}tre. Version fran\c{c}aise:
Analyse non standard et repr\'{e}sentation du r\'{e}el: deux exemples en
automatique, accessible sur {\tt
http$:$//hal.inria.fr/inria-00163365}.

\bibitem{mboup}M. Mboup, Parameter estimation via differential
algebra and operational calculus, Manuscrit, 2006. Accessible sur
{\tt http$:$//hal.inria.fr/inria-00138294}.

\bibitem{ath}M. Mboup, C. Join, M. Fliess, A revised look at numerical
differentiation with an application to nonlinear feedback control,
Proc. 15$^{th}$ Mediterrean Conf. Control Automation - MED'2007,
Ath\`{e}nes, 2007. Accessible sur {\tt
http$:$//hal.inria.fr/inria-00142588}.

\bibitem{McC}J. McConnell,  J. Robson,  Noncommutative Noetherian Rings, American
Mathematical Society, Providence, RI, 2000.

\bibitem{nelson}E. Nelson, Radically Elementary Probability Theory, Princeton
University Press, Princeton, NJ, 1987. Accessible sur {\tt
http$:$//www.math.princeton.edu/\%7Enelson/books/rept.pdf}.

\bibitem{neves}A. Neves, M.D. Miranda, M. Mboup, Algebraic parameter
estimation of damped exponentials, Proc. 15$^{th}$ Europ. Signal
Processing Conf. - EUSIPCO 2007, Pozna\'{n}, 2007. Accesible sur
{\tt http$:$//hal.inria.fr/inria-00179732}.

\bibitem{proakis0}J.G. Proakis, Digital Communications, $4^{th}$ ed.,
McGraw-Hill, New York, 2001.


\bibitem{proakis}J.G. Proakis, M. Salehi, Communication Systems
Engineering, $2^{nd}$ ed., Prentice Hall, Upper Saddle River, NJ,
2002.


\bibitem{singer}M. van der Put, M.F. Singer, Galois Theory of Linear
Differential Equations, Springer, Berlin, 2003.

\bibitem{robinson}A. Robinson, Non-Standard Analysis, $2^{nd}$ ed.,
North-Holland, Amsterdam, 1974.

\bibitem{bell}C.E. Shannon, A mathematical theory of communication,
Bell Syst. Tech. J., 27 (1948) 379-457 \& 623-656.

\bibitem{ire}C.E. Shannon, Communication in the presence of noise, Proc. IRE,
37 (1949) 10-21.

\bibitem{trapero}J.R. Trapero, H. Sira-Ram\'{\i}rez, V.F. Battle, An algebraic frequency
estimator for a biased and noisy sinusoidal signal, Signal
Processing, 87 (2007) 1188-1201.

\bibitem{trapero-bis}J.R. Trapero, H. Sira-Ram\'{\i}rez, V.F. Battle, A fast
on-line frequency estimator of lightly damped vibrations in flexible
structures, J. Sound Vibration, 307 (2007) 365-378.

\bibitem{yosida}K. Yosida, Operational Calculus$:$ A Theory of
Hyperfunctions, Springer, New York, 1984 (translated from the
Japanese).


\end{thebibliography}
\end{document}